\documentclass[prd,aps,superscriptaddress,amssymb,nofootinbib,a4paper]{revtex4}
\usepackage{latexsym}
\usepackage{epsfig,amsmath,amssymb}
\usepackage{hyperref}

\def\Re{\mathop{\rm Re}\nolimits}

\newcommand{\be}{\begin{equation}}
\newcommand{\ee}{\end{equation}}
\newcommand{\bea}{\begin{eqnarray}} 
\newcommand{\eea}{\end{eqnarray}}

\def\cM{{\cal M}} \def\cN{{\cal N}}  
 \def\cE{{\cal E}}   

\def\rmi{{\mathrm i}}
\newcommand{\U}{\mathop{\rm {}U}}
\newcommand{\SU}{\mathop{\rm SU}}
\newcommand{\ft}[2]{{\textstyle\frac{#1}{#2}}}

\begin{document}

\title{Gravitating cosmic strings with flat directions}

\author{Betti Hartmann}
\email{b.hartmann@jacobs-university.de}
\affiliation{School of Engineering and Science, Jacobs University Bremen,
28759 Bremen, Germany}

\author{Asier Lopez-Eiguren}
\email{asier.lopez@ehu.es}
\affiliation{Department of Theoretical Physics, University of the Basque Country UPV/EHU, 48080 Bilbao, Spain}

\author{Kepa Sousa}
\email{k.sousa@jacobs-university.de}
\affiliation{School of Engineering and Science, Jacobs University Bremen,
28759 Bremen, Germany}
\affiliation{Department of Theoretical Physics, University of the Basque Country UPV/EHU, 
48080 Bilbao, Spain}

\author{Jon Urrestilla}
\email{jon.urrestilla@ehu.es}
\affiliation{Department of Theoretical Physics, University of the Basque Country UPV/EHU, 48080 Bilbao, Spain}

\date{\today}

\begin{abstract}
We study field theoretical models for cosmic strings with flat directions in curved space-time. 
More precisely, we consider minimal models with 
semilocal, axionic and tachyonic strings, respectively. In flat space-time, isolated static and straight  cosmic strings solutions of these models have a flat direction, {\it i.e.}, 
a uniparametric family of configurations with the same energy exists which is associated
with a zero mode. We prove that this zero mode survives coupling to gravity, and 
study the role of the flat direction when coupling the string to 
gravity. Even though the total energy of the solution is the same, and thus  the global properties of the family of solutions 
 remains unchanged, 
the energy density, and therefore the gravitational properties, are different. The local structure of the solutions depends strongly
on the value of the parameter describing the flat direction; for example, for a supermassive string, 
the value of the free parameter can determine the size of the space-time.
\end{abstract}

\maketitle

\section{Introduction}

In several field theoretical models which allow for string-like
objects it has been observed that families of 
solutions which have different local behaviour of the fields appear, but possess
the same value of the energy per unit length. 
These zero modes do not disappear 
when coupling the models minimally to gravity. 
Therefore, despite the fact that the global properties of the family of solutions 
will remain unchanged,  the gravitational properties might change, since, after all, 
gravity is intrinsically locally defined.

The simplest model that  is frequently used to describe cosmic strings is the $U(1)$ Abelian--Higgs 
model. The models we study in this paper are somewhat more complicated than this one, but will be 
closely related to it since the solutions can (sometimes) be seen as embedded Abelian-Higgs strings. 
 Semilocal strings are solutions of an $SU(2)_{global}\times U(1)_{local}$ model which -- in fact --
corresponds to the Standard Model of Particle physics in the limit $\sin^2\theta_{\rm w}=1$, where
$\theta_{\rm w}$ is the Weinberg angle. The simplest semilocal string solution is an
embedded Abelian--Higgs solution \cite{Vachaspati:1991dz,Achucarro:1999it}. A detailed analysis of the stability
of these embedded solutions has shown \cite{Hindmarsh:1991jq,Hindmarsh:1992yy} that they are unstable (stable)
if the Higgs boson mass is larger (smaller) than the gauge boson mass. In the case
of equality of the two masses, the solutions fulfill a Bogomolnyi--Prasad--Sommerfield (BPS) \cite{Bogomolny:1975de}
bound such that their energy per unit length is directly proportional to the winding number.
Interestingly, it has been observed \cite{Hindmarsh:1991jq,Hindmarsh:1992yy} that in this BPS limit, it is possible to find a one-parameter family of single static and straight cosmic string solutions: the Goldstone field can form a non-vanishing condensate
inside the string core and the energy per unit length is independent of this value, which itself
is related to the width of the string.
These solutions are also sometimes denominated ``skyrmions'' and have been related to the 
zero mode associated with the width of the semilocal strings present
in the BPS limit. Semilocal strings have been studied in cosmological settings both in the context of  
their formation \cite{Achucarro:1992hs,Achucarro:1997cx,Achucarro:1998ux, Urrestilla:2001dd}, network properties \cite{Achucarro:2007sp} and their CMB implications \cite{Urrestilla:2007sf}.

The second model that we are investigating in this paper is a field theoretical 
model for so-called axionic and tachyonic strings.
This model, introduced in Blanco-Pillado \emph{et al.} \cite{BlancoPillado:2005xx}, originally 
describes an unstable D--brane anti--D--brane pair  
and allows for the existence of 
non-singular BPS strings. In fact, depending on the boundary conditions
imposed on the matter fields, the model accommodates 
three different types of strings \cite{Achucarro:2006ef}: $\phi$-strings (tachyonic), $s$-strings (axionic),
and a hybrid of both types. 
As in the semilocal model,  each of these types of strings has a family of solutions of the same energy parametrized by one 
parameter. For tachyonic and axionic strings, respectively, 
this parameter is associated with the ``width'' of the strings - very similar to what happens
in the semilocal model. On the other hand, for hybrid strings it measures the contribution
of the tachyonic string in relation to the axionic string.
These strings,  in particular the tachyonic strings, share some qualitative 
features with semilocal strings.

The gravitational properties of field theoretical cosmic string solutions have also
been discussed in the literature. The most thorough study has been done for  Abelian--Higgs strings 
by minimally coupling the Abelian--Higgs model to gravity \cite{Christensen:1999wb,Brihaye:2000qr}. 
Far away from the 
core of the string, the space-time has a deficit angle, i.e. corresponds to
Minkowski space-time minus a wedge \cite{Garfinkle:1985hr}. The deficit angle is  proportional to the energy per unit length of the string. 
If the vacuum expectation value (vev) of the Higgs field is sufficiently large
(corresponding to very heavy strings that have formed at energies bigger than the GUT scale), 
the deficit angle becomes larger than $2\pi$. These solutions
are the so-called ``supermassive strings'' studied in \cite{Laguna:1989rx,Ortiz:1990tn} and possess
a singularity at a maximal value of the radial coordinate, at which the
angular part of the metric vanishes.
Gravitating semilocal strings have first been 
studied in \cite{Hartmann:2009tc}.

In this paper we reinvestigate the gravitating semilocal strings and point out further details.
Moreover, we investigate gravitating axionic and tachyonic strings 
focusing in all cases on the role of the zero mode. Our paper is organised as follows: in Section \ref{model}, 
we describe the models and give the equations of motion. In order to simplify the calculations we will use a 
standard result in the  context of supergravity theories: cosmic string configurations preserving a fraction of 
the supersymmetries in a supergravity theory saturate a BPS bound, and  the corresponding Einstein 
equations admit a first integral, 
the so called \emph{gravitino equation}, (see for example \cite{Dvali:2003zh}). The models we will discuss here can 
be embedded in $\mathcal{N}=1$ supergravity, \cite{Dasgupta:2004dw,BlancoPillado:2005xx}, and moreover 
the corresponding cosmic string solutions leave unbroken half of the supersymmetries of the original theory. 
Here, following the results in \cite{Gibbons:1992gt}, we will rederive the first integrals of the Einstein equations for the present models treating them as non-supersymmetric theories, and therefore, without making any explicit reference to supersymmetry. The rest of the field equations will be obtained using a BPS-type of argument valid for static cylindrically symmetric configurations  in a gravitational theory, that is, searching for the conditions required to minimize an appropriate energy functional.  The resulting equations will turn out to be a set of first order differential equations for both models analog to those found in flat space-time \cite{Achucarro:1999it,BlancoPillado:2005xx}, and in particular we   recover the results already found in \cite{Hartmann:2009tc} for the semilocal case.
    
 In section \ref{NR} we discuss the numerical results and conclude in Section 
\ref{conclusions}.

\section{The models and their flat directions}
\label{model}

The models we are studying have an action of the  following form 
\begin{equation}
\label{action}
S=-  \frac{1}{2} M_p^{2} \int  d^4 x \, \sqrt{- \mathrm{det} \, g} \, R +  \int   d^4 x  \,
 \sqrt{- \mathrm{det} \, g} \,{\cal L}_{m}  \ , 
\end{equation}
where the gravitational strength coupling is given in term of the  reduced Plank mass $M_p^{-2} = 8 \pi G$.
The space-time metric has  signature $(-,+,+,+)$.  The  Ricci tensor scalar follows conventions from \cite{Dvali:2003zh}.

We will study two types of matter Lagrangians
${\cal L}_{m}$, the semilocal model \cite{Vachaspati:1991dz,Achucarro:1999it}, and the axionic D-term model presented in \cite{BlancoPillado:2005xx}. 
First we will discuss the gravity part of the model and then we will describe the matter content in the 
subsections below. In particular, we will point out what type of string-like solutions these models can 
accommodate and discuss the role that gravity plays in connection to the zero modes.\\

In the present work we will focus on static cylindrically symmetric configurations invariant under boosts 
along the axis of symmetry which, without loss of  generality,  we can take to be the  $z-$axis.  The most 
general  line element consistent with these symmetries is:
\begin{equation}
ds^2=- N^2(r)dt^2+dr^2+L^2(r)d\varphi^2+N^2(r)dz^2 \ ,
\label{metric}
\end{equation}
where $(r,\varphi,z)$ are cylindrical coordinates. The non-vanishing components of the Ricci tensor 
$R_{\mu}^{\nu}$ then read \cite{Christensen:1999wb}:
\begin{eqnarray}
R_0^0=\frac{(LNN')'}{N^2 L} \ \ , \ \ R_{r}^{r} = \frac{2N''}{N}+\frac{L''}{L} \ \ \ , \ 
\ R_{\varphi}^{\varphi}= \frac{(N^2 L')'}{N^2 L} \ \ \ , \ \ \ R_z^z=R_0^0  \ ,
\end{eqnarray}
where the prime denotes the derivative with respect to $r$. With our conventions  the Einstein equations 
have the following form

\begin{equation}
\label{einstein}
 R_{\mu\nu}= - 
M_p^{-2} \left(T_{\mu\nu} - \frac{1}{2} g_{\mu\nu} T\right) \ ,
\end{equation}
where $T=T^{\sigma}_{\sigma}$ is the trace of the energy-momentum tensor which is given by
\begin{equation}
 T_{\mu\nu}=-2\frac{\partial {\cal L}_{m}}{\partial g^{\mu\nu}} + g_{\mu\nu} {\cal L}_m  \ .
\end{equation}
In  \cite{Gibbons:1992gt} it was shown that the energy momentum tensor  of the semilocal model has a very 
simple form  for field configurations  satisfying the BPS bound. In particular, the only non-vanishing 
components of $T_{\mu}^\nu$ are proportional to the field strength of an auxiliary vector field  $A_\mu^B$, 
which is a function of the gauge boson and the scalar fields involved in the string configuration:
\be
T_r^r = T_\varphi^\varphi=0, \qquad T_t^t = T_z^z  =
\pm L^{-1} (\partial_r A_\varphi^B - \partial_\varphi  A_r^B) , 
\label{simpleT}
\ee 
As we shall see below this statement   is also true for the axionic D-term string model proposed in \cite{BlancoPillado:2005xx}.  \\

If the energy-momentum tensor is of the form (\ref{simpleT})  the metric function $N(r)$  becomes a 
constant \cite{VilShe94,Hindmarsh:1994re}, and thus choosing the coordinates conveniently, we can always set it to 
one,  $N(r)\equiv 1$. To obtain the equation of motion for the remaining metric function $L(r)$ we use the 
$\varphi\varphi$-component of (\ref{einstein}) which reads
\begin{equation}
\frac{L''}{L}=-M_p^{-2} \left(T^{\varphi}_{\varphi} - \frac{1}{2} T\right)= \frac{1}{2}M_p^{-2} T\ .
\end{equation}
In order for the metric to be regular at the $z-$axis we need to  impose the following  boundary 
conditions
\begin{equation}
 L(0)=0 \ \ , \  \ L'(0)=1 \ .
\label{metricBC}
\end{equation}

For cylindrically symmetric field configurations it is possible to choose a gauge where the vector field 
$A_\mu^B$ satisfies  $A_r^B =0$ and $A_\varphi^B =A_\varphi^B(r)$ and therefore, as long as the BPS 
bound is saturated, we can find a {\bf first integral} to the Einstein equations\footnote{In the context of 
supergravity theories this equation is know as the \emph{gravitino equation} \cite{Dvali:2003zh}.}:
\be
 L''= \pm M_p^{-2} \, (A_\varphi^{B})' \qquad \Longrightarrow \qquad L'= 1 \pm M_p^{-2} \,A_\varphi^B.
\label{gravitinoEq}
\ee  
Here the integration constant is fixed by the boundary conditions (\ref{metricBC}), since regularity also 
requires that $A_\varphi^B(0)=0$.  This first integral of the Einstein equations allows us to express the 
deficit angle of the string configurations  in terms  of the asymptotic value of  the  vector field $A_\mu^B$  
\cite{VilShe94} 
\be
\delta = 2 \pi (1- L'|_{r=\infty}) = \mp 2 \pi \, M_p^{-2}   \,A_\varphi^B|_{r=\infty}.
\label{deficit}
\ee
In order to discuss the existence of zero-modes in these theories we need an appropriate definition of 
the energy which is valid for general curved space-times.  We use  the same definition as in \cite{Dvali:2003zh}, 
which is valid for  time independent configurations,  and is obtained adding a  Gibbons-Hawking term to 
the action:
 \be
E =  - S - S_{GH}, \qquad S_{GH} =  M_p^{2}  \int_{\partial \cM}  \sqrt{-\mathrm{det}\, \tilde g} \; K.
\label{energyDef}
\ee
Here $K$ is the trace of the second fundamental form of the metric $g_{\mu\nu}$ at the space-time 
boundary $\partial \cM$, and  $\tilde g_{\alpha \beta}$  is the metric induced at $\partial \cM$.  Note that field 
configurations minimizing this energy functional also extremize the action, and therefore are solutions to the equations
of motion \cite{Gibbons:1976ue}. In order to obtain a finite value for the energy we calculate the integrals in (\ref{energyDef}) only over the
 plane orthogonal to the string, i.e. over the $r$ and $\varphi$ coordinates. Then, choosing  the boundary
 to be a cylinder centered on the $z-$axis with an arbitrary  large radius $r \to \infty$ we have (after setting
 $N(r)=1$)
\be
\sqrt{-\mathrm{det}\, \tilde g} \; K = L'(r),
\ee
and thus the Gibbons-Hawking term can be written in terms of the deficit angle
\be
 S_{GH} = M_p^{2}  \int_{\partial \cM}  \sqrt{-\mathrm{det}\, \tilde g} \; K =  
2\pi \, M_p^{2}    \left( L'|_{r=\infty} -L'|_{r=0}  \right)  = - M_p^{2} \, \delta.
\ee
Due to the Einstein equations, the action $S$ vanishes on shell; and  we find the following relation between the energy of  the string 
and its deficit angle:
\be
E = M_p^{2} \, \delta = \mp 2 \pi \, A_\varphi^B|_{r=\infty} ,
\label{energyDeficit}
\ee 
where the second equality can be obtained using equation (\ref{deficit}). In the following subsections we 
will arrive to the same result following a Bogomolnyi type of argument \cite{Bogomolny:1975de} which involves 
rewriting the energy functional as a sum of positive terms plus a boundary term. 

It is possible to show that for field configurations with the same symmetries as the ones we discuss here  the definition of the energy (\ref{energyDef}) agrees with the one used in \cite{Hartmann:2009tc},
\be
E = \int  d^2 x \, \sqrt{- \mathrm{det} \, g} \, T_0^0\,.
\ee
In order to check this is sufficient to note that using the ansatz for the metric (\ref{metric}), the Einstein-Hilbert term becomes a boundary term which is exactly cancelled by $S_{GH}$, and that any contribution to the action involving time derivatives must be zero. 

\subsection{Semilocal strings}

First, we will reconsider the semilocal model which possesses a  $SU(2)_{\rm global} \times U(1)_{\rm local}$ 
symmetry \cite{Vachaspati:1991dz,Achucarro:1999it} and is given by the following matter Lagrangian density
\begin{equation}
{\cal L}^{\rm SL}_m=- \left(D_{\mu} \Phi\right)^{\dagger} D^{\mu} \Phi -\frac{1}{4 g^2} 
F_{\mu\nu} F^{\mu\nu} -\frac{\lambda}{2}\left(\xi - \Phi^{\dagger} \Phi \right)^2   \ ,
\label{sm}
\end{equation}
where $F_{\mu\nu}=\partial_{\mu} A_{\mu} - \partial_{\nu} A_{\mu}$ is the field strength tensor of the $U(1)$ 
gauge field and $D_{\mu} \Phi=(\partial_{\mu} - i A_{\mu}) \Phi$ is the  covariant derivative of the complex 
scalar field doublet $\Phi=(\phi_1,\phi_2)^T$. The constant  $g$ denotes the gauge coupling, $\lambda$ 
the self-coupling of the scalar fields and $\sqrt{\xi}$ is their vacuum expectation value. \\

If the couplings satisfy the  Bogomolnyi limit $g = \sqrt{\lambda}$, and working in the gauge $A_0= A_z =0$, the energy functional  (\ref{energyDef})  for the stationary string with a translational symmetry along the $z-$axis can be given by
\bea
\label{slener1}
E &=& \int dr d\varphi \, L(r) \left[\left(D_{r} \Phi\right)^{\dagger} D_{r} \Phi+L^{-2}\left( D_{\varphi} \Phi\right)^{\dagger} D_{\varphi} \Phi +\frac{1}{2 g^2} 
L^{-2} F_{r\varphi} F_{r\varphi} +\frac{g^2}{2}\left(\xi - \Phi^{\dagger} \Phi \right)^2  \right]\,,\eea
 which can be rearranged  in the following way:
\bea
E &=& \int dr d\varphi \, L(r) \Big[ (D_r \Phi \pm \rmi L^{-1} D_\varphi  \Phi)^\dag 
(D_r \Phi \pm \rmi  L^{-1} D_\varphi \Phi) + \nonumber \\ 
 &&\frac{1}{2 g^2}\left( L^{-1} F_{r \varphi}  \mp g^2 (\xi -  \Phi^\dag \Phi )\right)^2  \Big]
 \mp   \int dr d\varphi F_{r \varphi}^B,
\label{BogAction}
\eea
where we have   used the explicit form of the  metric (\ref{metric}) after setting $N(r)=1$, and we have 
introduced the auxiliary vector field $A_\mu^B$  and its field strength\footnote{In the context of $\cN=1$ 
supergravity, $A_\mu^B$ is known as the gravitino $\U(1)$ connection.}: 
\be
A_\mu^B  \equiv \ft{\rmi}{2} (\Phi^\dag D_\mu \Phi - D_\mu \Phi^\dag\Phi  ) -  
\xi \, A_\mu, \qquad F_{\mu \nu}^B \equiv \partial_\mu A_\nu^B -\partial_\nu A_\mu^B.
\label{SLconnection}
\ee
  In order to obtain finite energy configurations (\ref{slener1}) we have to require that far away from the center of the string
 the fields are in the vacuum, and that the covariant derivatives of the fields vanish
\be
\Phi^\dag \Phi|_{r\to \infty} = \xi \ , \qquad  D_\mu \Phi|_{r\to \infty} =0 \ .
\label{semilocaBC}
\ee
Therefore, using Stokes' theorem, we find the following lower bound for the energy of the cosmic string
\be
E  \ge \mp \int dr d\varphi \,    F_{r \varphi}^B  =  \pm  \xi  \int dr d\varphi \, F_{r \varphi},
\label{energybound}
\ee 
where the last integral is the total magnetic flux trapped inside the string.  Any field configuration which saturates this bound is at a local minimum of the energy functional and thus, as we argued in the previous subsection, it must be a solution to the equations of motion. In particular, the bound is saturated by cosmic string configurations which satisfy the BPS equations:
\be
D_r \Phi \pm \rmi L^{-1}  D_\varphi  \Phi =  0, \qquad 
L ^{-1} \,F_{r \varphi}  \mp g^2  (\xi -  \Phi^\dag \Phi ) = 0. 
\label{SLBPSeqs}
\ee

We use the following ansatz for a static straight cosmic string lying along the $z-$axis 
\cite{Vachaspati:1991dz,Achucarro:1999it}:
\begin{equation}
\phi_1= \sqrt{\xi} f(r) e^{i n\varphi}  \  , \qquad 
\phi_2=\sqrt{\xi} h(r) e^{i m \varphi} \  ,   \qquad  A_{\varphi}= \pm v(r),
\label{SLansatz}
\end{equation}
with all the other components of the gauge field set to zero. Here $n$ and $m$ are the winding numbers 
of the fields $\phi_1$ and $\phi_2$ respectively and,  without loss of generality, we will 
assume\footnote{The case $n=m$ can  always be rotated into a Nielsen-Olesen string using a $\SU(2)$ 
transformation, and therefore  will not discuss it. } $|n|>|m|$. By using the rescalings 
$r\rightarrow r/(g\sqrt{\xi})$ and $L(r)\rightarrow L(r)/(g\sqrt{\xi})$ the BPS equations read
\begin{eqnarray}
\label{fieldeq1}
 f' + \frac{(v-|n|)}{L} f =  0  \  ,  \qquad 
    h' + \frac{(v-|m|)}{L} h =  0  \  , \qquad
\frac{v'}{L} + (f^2 +h^2 -1) =  0,  
\label{BPSprofiles}
\end{eqnarray}
which have to be solved subject to appropriate boundary conditions that result from 
imposing (\ref{semilocaBC}) and regularity at the origin  \cite{Vachaspati:1991dz,Achucarro:1999it}
\bea
& & f_{\infty}^2+h_{\infty}^2= 1  \ , \qquad    v(r \to \infty) = |n| \ ,    \label{bc_semilocal1} \\
&& f(0)=0 \  , \qquad  v(0)=0 \ , \qquad  h'_{m=0}(0)=0  \; \textrm{ or } \; h_{m\neq0}(0)=0 \ ,  
\label{bc_semilocal2}
\eea
where $f_{\infty}=f(r=\infty)$ and $h_{\infty}=h(r=\infty)$.
The choice of signs of the winding numbers $n$ and $m$ in (\ref{BPSprofiles}) ensures that these 
conditions can be met. Indeed, the BPS equations  (\ref{SLBPSeqs}) with the upper sign can only 
be solved provided $n,m>0$, while the lower sign requires $n,m<0$.\\

The first two BPS equations (\ref{BPSprofiles})  imply  that  the profile functions $f$ and $h$  must be 
related to each other \cite{Vachaspati:1991dz,Achucarro:1999it}
\be
\log h = \log f - ( |n| - |m| )  \int \frac{dr}{L(r)} + \kappa  \qquad  \Longrightarrow
 \qquad  h =  c \cdot f \cdot  \exp \left( (|m|- |n|) \int \frac{dr}{L(r)} \right) 
\label{family}
\ee
and therefore, there is a one-parameter family of solutions characterized by a real constant 
$\kappa = \log c$ . As in flat space, this family is degenerate in energy, i.e. the total energy does 
not depend on the parameter  $\kappa$. This can be checked inserting the ansatz for the gauge boson 
(\ref{SLansatz}) and its boundary conditions in (\ref{energybound}):
\be
 E = \pm \xi \int d \varphi \,  A_\varphi|_{r= \infty} =  2 \pi |n|  \xi \ .   
\label{SLstringenergy}
\ee
For field configurations safisfying the BPS equations (\ref{SLBPSeqs}) it is easy to check that the energy momentum tensor takes the form (\ref{simpleT}), with $A_\mu^B$ given by (\ref{SLconnection}). Then, the profile function of the metric $L(r)$ can be obtained from  (\ref{gravitinoEq}), which using the ansatz 
(\ref{SLansatz})  takes the form
\be
 L'=1  - \alpha^2 \, \left( (|n|-v) f^2  +  (|m|- v) h^2 + v \right),
\ee
where $\alpha \equiv  M_p^{-1} \sqrt{\xi}$  is the vacuum expectation value of $\phi$ measured in Planck 
masses. From the previous  two equations it is possible to recover the relation (\ref{energyDeficit}) between
 the   deficit angle of the BPS strings and their energy 
\be
\delta = M_p^{-2}\, E = 2 \pi |n| \alpha^2  \ . 
\ee
Note that the choice of signs  made above for the winding numbers ensures that the energy, and thus 
the deficit angle, are positive.

\subsection{Axionic and tachyonic strings}

In this subsection we are going to consider the cosmic string solutions of the \emph{axionic $D-$term 
model} studied in flat space-time  in \cite{BlancoPillado:2005xx,Achucarro:2006ef}. The model describes the dynamics of two complex 
scalar fields,  the tachyon $\phi$, and an axio-dilaton $S=s+\rmi a$ ($s>0$), coupled to a  $U(1)$ gauge field
 $A_{\mu}$. The lagrangian density reads
\begin{equation}
\mathcal{L}^{\rm A}_m=- D_\mu \bar \phi  D^{\mu} \phi - K_{S \bar S} D_{\mu} \bar S
 D^{\mu} S - 
\frac{1}{4 g^2} F^{\mu \nu } F_{\mu \nu} -\frac{g^2}{2}   \left( \xi + 2 \delta K_S - q \bar \phi \phi\right)^2   
\label{axionAction}
\end{equation}
where $F_{\mu \nu}= F_{\mu\nu}=\partial_{\mu} A_{\nu} - \partial_{\nu} A_{\mu}$ is the $U(1)$ field strength, 
and  the covariant derivatives of the tachyon and the axio-dilaton are given by 
$D_{\mu} \phi=(\partial_{\mu} - i q A_{\mu}) \phi$ and $D_\mu S=\partial_{\mu} S +i 2 \delta  A_{\mu}$ 
respectively. The model depends on three continuous parameters: the gauge coupling $g$, the charge of 
the axio-dilaton $\delta$, and $\xi$ which determines the expectation value of the fields. The constant $q$ 
is an integer which represents the $\U(1)$ charge of the tachyon. The lagrangian density also involves the 
derivatives of the K\"ahler potential associated with the  axio-dilaton field $K(S, \bar S)$,  which is a real function 
chosen to have the following form: 
\begin{equation}
 K(S, \bar S)= -M_p^2 \log (S+\bar{S}).
\end{equation}
The quantities $K_S$ and $K_{S\bar{S}}$ denote  the derivatives of the K\"ahler potential with respect  to 
the fields $S$ and $\bar{S}$,  and they are given by
\begin{equation}
K_S=-M_p^2 \frac{1}{S+\bar{S}}   \ , \  \qquad K_{S\bar{S}}=M_p^2 \frac{1}{(S+\bar{S})^2}   \ .
\end{equation}
This lagrangian density is the bosonic part of a supersymmetric model with a $D-$term scalar potential, and 
therefore the couplings satisfy the Bogomolnyi limit \cite{Davis:1997bs}. Thus, without imposing further constraints 
in the couplings, the energy of a static configuration 
\be
E = \int dr d \varphi \, L(r) \left(D_r \bar \phi  D_{r} \phi+L^{-2}D_\varphi \bar \phi  D_{\varphi} \phi + K_{S \bar S} D_{\mu} \bar S
 D^{\mu} S + 
\frac{1}{2 g^2} L^{-2}F_{r \varphi } F_{r \varphi} +\frac{g^2}{2}   \left( \xi + 2 \delta K_S - q \bar \phi \phi\right)^2\right)\ ,
\label{AxioDilatonEnergy}
\ee
can be written  in the Bogomolnyi
form as follows
\bea
E &=& \int dr d \varphi \, L(r) \Big[|D_r \phi \pm \rmi L^{-1} D_\varphi  \phi |^2 + 
K_{S\bar{S}}  | D_r  S \pm \rmi  L^{-1} D_\varphi S|^2 \nonumber \\
 &&+ \frac{1}{2g^{2}}\left( \, L^{-1} F_{r \varphi}  \mp g^{2} (\xi+ 2 \delta K_S- q |\phi|^2)  \right)^2 \Big]  
\mp  \int dr d\varphi F_{r \varphi}^B \ ,
\label{Bog1Action}
\eea
where we have defined the composite vector field $A_\mu^B$ and its field strength
\be
A_\mu^B \equiv  \ft{\rmi}{2} (\bar \phi D_\mu \phi - \phi D_\mu \bar \phi   ) -
  \frac{\rmi M_p^2}{4 \, \Re  S}   ( D_\mu S - D_\mu \bar S )  - \xi A_\mu,
 \qquad F_{\mu \nu}^B \equiv \partial_\nu A_\mu^B -  \partial_\mu A_\mu^B.
\ee
As in the semilocal model, if we restrict ourselves to finite energy configurations, we have to impose  
that far away from the string core the fields are in the vacuum, and that the covariant derivatives vanish
\be
(q |\phi|^2 - 2 \delta K_S)_{r\to \infty}= \xi   \ , \qquad D_\mu \phi|_{r\to \infty} \ , 
\qquad D_\mu S|_{r\to \infty} =0 \ .
\ee
With these boundary conditions we find a lower bound for the energy similar to  (\ref{energybound}), and 
the corresponding BPS equations (which ensure that  the bound is saturated) read 
\begin{eqnarray}
(D_r \pm i L^{-1}D_\varphi)\phi = 0, \qquad
(  D_r \pm i L^{-1} D_\varphi ) S =0, \qquad 
 L^{-1} \, F_{r \varphi} \mp g^2 (\xi+2 \delta K_S- q |\phi|^2) =0. 
\label{bogol}
\end{eqnarray} 
As we mentioned earlier in the paper, the energy momentum tensor reduces to a very simple form 
for field configurations which satisfy the previous equations. Indeed,  the only non-vanishing components 
are given by 
\be
T_{t}^t = T_{z}^z = \pm L^{-1} F_{r\varphi}^B, 
\ee
and therefore as we discussed before the Einstein equations admit the first integral  (\ref{gravitinoEq}).\\

In order to solve the BPS equations we use the ansatz for the matter fields proposed in  \cite{BlancoPillado:2005xx},
 which represents a  static straight cosmic string along the $z-$axis: 
\begin{eqnarray}
\phi =\sqrt{\xi/ q} \,  f(r) e^{i n \varphi} & \qquad s^{-1}(r) = \xi / ( \delta M_p^2) \, h(r)^2 \nonumber \\
a=2 \delta \, m \varphi  \qquad & \qquad A_\varphi = \pm v(r) \ .
\label{ansatz}
\end{eqnarray}
 After rescaling the radial coordinate and the metric profile function, $r\rightarrow  r/(g\sqrt{\xi})$,  
$L \rightarrow L/( g\sqrt{\xi}) $, the BPS equations become:
\begin{equation}
  f^{\prime}+ \frac{(qv-|n|)}{L}f=0, \qquad h^{\prime}+ \alpha^2 q \frac{(v-|m|)}{L}h^3=0, 
\qquad \frac{v^{\prime}}{L} + (f^2+h^2-1)=0,
\label{axionicBPSeq}
\end{equation}
where $\alpha=M_p^{-1}  \sqrt{\xi/q}$. The signs of the winding numbers $n$ and $m$  are fixed by requiring  $f(r)$ to be regular and
 $h(r)^2>0$ for $r\to 0$. As in the semilocal case, we can use the first two equations to find  a relation 
between the tachyon and the dilaton field
\begin{equation}
\frac{1}{\left(\alpha h\right)^2}=2\left(|n|-q|m|\right) \int \frac{dr}{L(r)}  -2\log f+\kappa \ ,
\end{equation}
and therefore the solutions are parametrized by an arbitrary constant  $\kappa$. Actually, this model admits 
three different families of cosmic string solutions depending on the boundary conditions at $r \to \infty$:

\begin{itemize}

\item $\phi$-strings (tachyonic)

In this type of strings the magnetic flux trapped in the core is induced by the winding of the tachyon field, 
which must satisfy $|n|>q|m|$ in order to solve the BPS equations.   The profile functions have the following
 asymptotic behavior 
\begin{equation}
\label{bc_tachyonic1}
f(r\rightarrow\infty)\rightarrow 1\,,\qquad h(r\rightarrow\infty)\rightarrow 0\,,\qquad 
v(r\rightarrow\infty)\rightarrow |n|/q
\end{equation}
so that the tachyon field acquires a non-vanishing expectation value far away  from the core while the 
function $h(r)$ tends to zero. For the solutions to be regular we also have to impose the following boundary
 conditions at the core of the strings $r\to 0$:
\be
\label{bc_tachyonic2}
 f(0)=0, \qquad  v(0)=0,  \qquad  h'_{m=0}(0)=0,  \quad \textrm{ or } \quad h_{m\neq0}(0)=0.
\ee
In the present work we will only consider the case $m=0$, where the profile function of the axion-dilaton,
 $h(r)$, can be non-zero at the core  creating a ``condensate''. Actually, as in the flat space-time analysis 
\cite{BlancoPillado:2005xx}, the value of $h(r=0)$ will be a  free parameter related to $\kappa$, which determines the 
width of the string. This family of solutions is degenerate in energy  
\be
 E = \pm \xi \int d \varphi \,  A_\varphi|_{r= \infty} =  2 \pi |n|\frac{ \xi}{q} \ , 
\label{TachyonEnergy}
\ee
and therefore the zero-mode survives the coupling to gravity. For simplicity, in our numerical analysis we will set the constant $q=1$.

\item $s$-strings (axionic)

In this case the magnetic flux inside the strings is induced by the winding of the axio-dilaton, $S$.  
The behaviour at infinity is
\begin{equation}
\label{bc_axionic1}
f(r\rightarrow\infty)\rightarrow 0\,,\qquad h(r\rightarrow\infty)\rightarrow 1\,,\qquad v(r\rightarrow
\infty)\rightarrow |m|  \ .
\end{equation}
It is now the dilaton which acquires a non-zero expectation value far from the core, while the tachyonic field
 tends to zero far from the core. Thus, the role of the  tachyonic and the dilatonic field is exchanged. These 
strings are solutions to the BPS  equations provided $|n|<q \, |m|$. Regularity at the origin imposes the 
following  boundary conditions
\be
\label{bc_axionic2}
h(0)=0, \qquad v(0)=0, \qquad f'_{n=0}(0)=0,  \quad \textrm{ or } \quad f_{n\neq0}(0)=0.
\ee
In the next section we discuss the properties of this family of solutions for the case $n=0$, where the 
value of the profile function of the tachyon at the center of the string, $f(r=0)$, is a free parameter related 
to $\kappa$. Again, this family of solutions is degenerate in energy
\be
 E = \pm \xi \int d \varphi \,  A_\varphi|_{r= \infty} =  2 \pi |m| \xi , 
\ee
 and thus the zero-mode still exists after coupling the model to gravity. Note that $s-$strings have a tension
 $q$ times larger than $\phi-$strings.  As for the tachyonic strings, we will restrict the analysis to the case $q=1$.

\item Hybrid-strings

In this case, both the tachyon and the dilaton field contribute to cancel the   scalar potential far away from 
the core, i.e., they both acquire a finite vacuum expectation value for $r \to \infty$. This can only happen 
provided the windings satisfy $|n|=q \, |m|$. Thus, in this case  we have the following boundary conditions 
at infinity
\begin{equation}
\label{bc_hybrid1}
f_{\infty}^2+h_{\infty}^2 = 1\,,\qquad v(r\rightarrow\infty)
= \frac{|n|}{q}=|m|  \ ,
\end{equation}
 where $f(r\rightarrow \infty)\equiv f_{\infty}$ and $h(r\rightarrow\infty)\equiv h_{\infty}$. Provided the 
previous constraints are satisfied, the value of $f_{\infty}$ is a free parameter, which will be again related to
$\kappa$. In this case $\kappa$ is not a measure of the width of the string,  instead, it is related to the 
relative contributions of the tachyonic and the axionic string to the tension of the string . The 
corresponding boundary conditions at the core of the string are given by
 \be
\label{bc_hybrid2}
 f(0)=0, \qquad  v(0)=0,  \qquad  h(0)=0.
\ee

The zeromode associated with the parameter $\kappa$ is not normalizable and thus we will not discuss it any further in the present work. Indeed, if we promote the parameter $\kappa$ to be time dependent  the corresponding effective  action for the zeromode gets a 
quadratically divergent contribution $\Lambda^2$, where $\Lambda$ is a cutoff which, in a cosmological setting, could be given by the distance to the closest cosmic string.  
\end{itemize}

Finally, the form of the equation for the metric profile function (\ref{gravitinoEq}) can be found using 
the ansatz (\ref{ansatz}):
\begin{equation}
L' = 1-\alpha^2\left[(|n|-q v) f^2 + q  (|m|-v) h^2 +q v \right].
\end{equation}
Similarly to the semilocal case, from the boundary conditions discussed above, it is straightforward to 
find a relation between the deficit angle of these strings and their energy: $\delta =M_p^{-2} q \, E$.\\

\section{Numerical results}
\label{NR}

We have solved the BPS equations  for both models, subject to appropriate boundary conditions, focusing on the interplay between the flat direction and gravity, in order words, we studied how the local field configuration changed the energy density and the properties of the solutions, even though the global energy was unchanged.

\subsection{Semilocal strings}

The gravitating properties of semilocal strings were first studied in \cite{Hartmann:2009tc}. Besides completing that analysis, 
we include this case in the present work because it shares several properties with the other cases to be studied.

\begin{figure}[!htb]
\includegraphics[width=6cm]{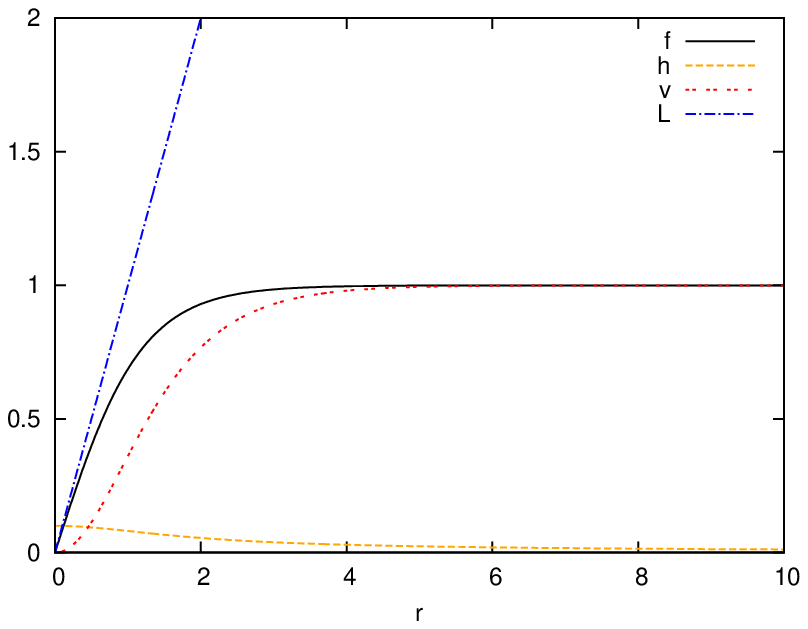}
\includegraphics[width=6cm]{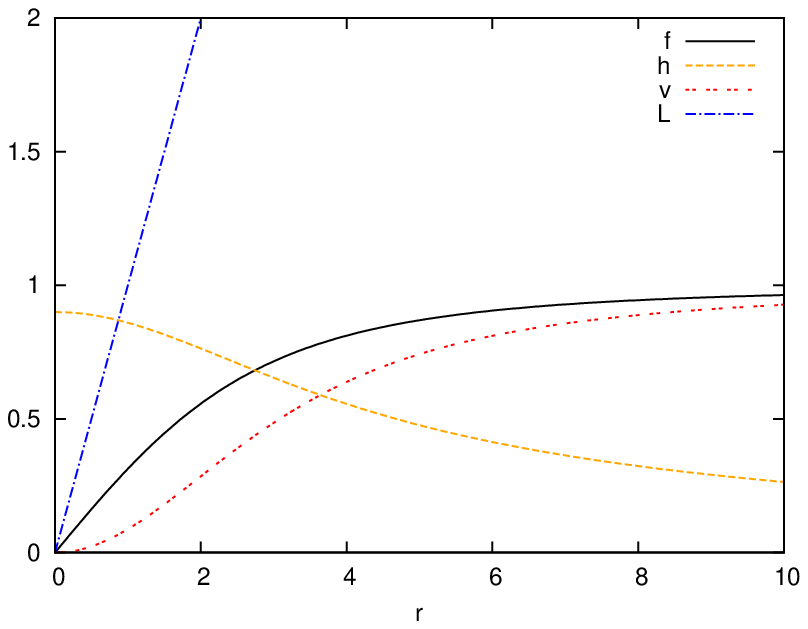} 
\caption{\label{fig:semi1}Profiles of a typical $n=1$ semilocal string solution uncoupled to gravity, with different values of the free parameter: $h(0)=0.1$ (left) and
$h(0)=0.9$ (right).}
\end{figure}

\begin{figure}[!htb]
\includegraphics[width=6cm]{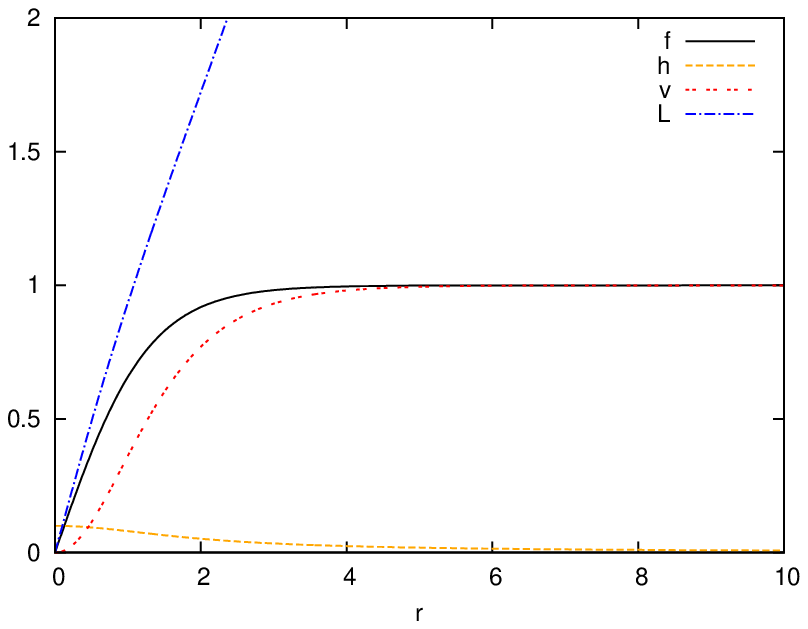}
\includegraphics[width=6cm]{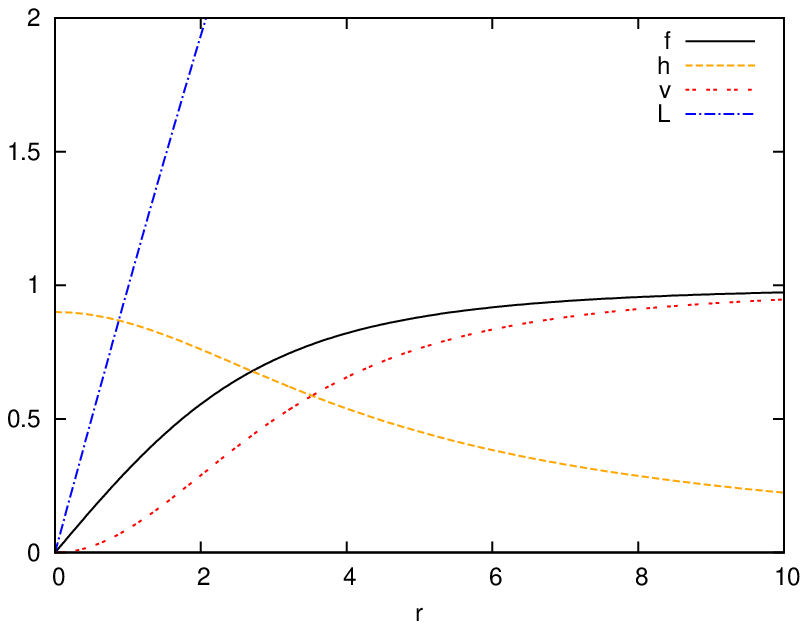} 
\caption{\label{fig:semi2}The profiles of a  gravitating $n=1$ semilocal string ($\alpha=0.5$) with the value of the free parameter $h(0)=0.1$ (left) and
$h(0)=0.9$ (right).}
\end{figure}

Fig.~\ref{fig:semi1} shows
two typical $n=1$ cosmic string solutions to this model, when the coupling to gravity is not included. The metric function $L\equiv r$ in this case. 
 These figures illustrate the effect of varying the value of the condensate $h$ at the center of the string $r=0$, i.e. the free parameter of the family of degenerate solutions.  For larger values of $h(r=0)$ the profile functions reach their asymptotic values farther away from the center of the string and thus, as we anticipated in the previous subsection, the width of the string core increases. 

 The form of the profiles when coupling to gravity, can be seen in Fig.~\ref{fig:semi2}.    Close to the core, the  profile functions have the same form as in a Minkowski background:
\be
L(r) \approx r + \ldots \, , \qquad
f(r) \approx f_0 \, r^{|n|} + \ldots \, , \qquad 
h(r) \approx h_0 \,  r^{|m|} + \ldots \, , \qquad
v(r) \approx \frac{1}{2} r^{2} + \ldots \, .
\label{semilocalAtZero}
\ee
However, far away from the core,  $r\to \infty$,  the metric takes the form
\be
ds^2 \approx - dt^2 + dr^2 + (1-|n|\alpha^2)^2 \, r^2 d \varphi^2 + dz^2 ,
\ee
 which corresponds to a conical space-time with a deficit angle $\delta = 2 \pi |n| \alpha^2$, and the behavior of the profile functions in the same limit is given by the following  asymptotic expansion:
\be
f(r) \approx 1 - \frac{1}{2} \left( \frac{r}{r_0} \right)^{\frac{2(|m|-|n|)}{1 - |n|\alpha^2}} + \ldots \, ,\qquad 
h(r) \approx \left( \frac{r}{r_0} \right)^{\frac{|m|-|n|}{1 - |n|\alpha^2}} + \ldots \, ,\qquad
v(r) \approx |n| - (|n|-|m|)\left( \frac{r}{r_0} \right)^{\frac{2(|m|-|n|)}{1 - |n|\alpha^2}} + \ldots \, ,
\label{semilocalAtInfty}
\ee
where $r_0$ is a parameter related to $h_0$ which determines the width of the string core. Setting the constant $\alpha=0$ we recover the corresponding expansions in flat space.

\begin{figure}[!htb]
\includegraphics[width=6cm]{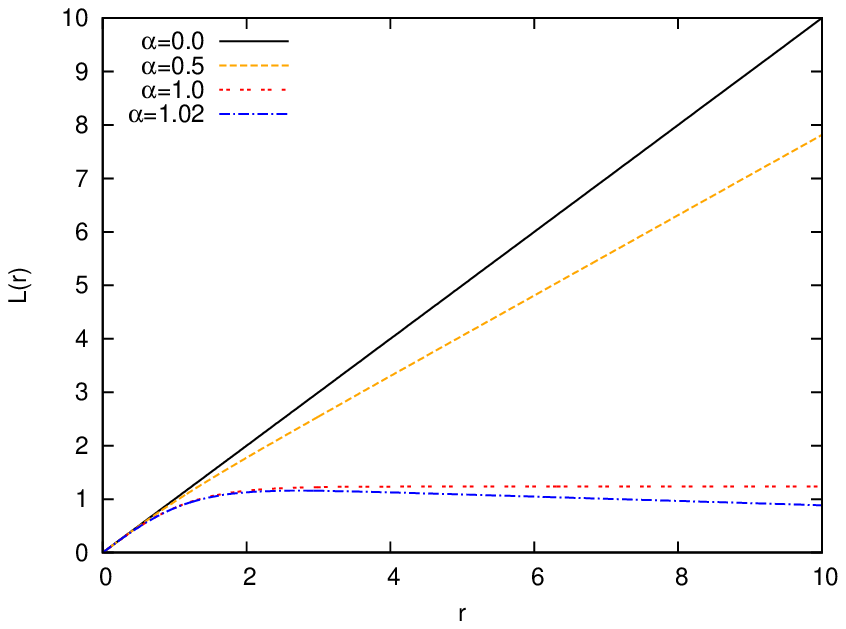} 
\includegraphics[width=6cm]{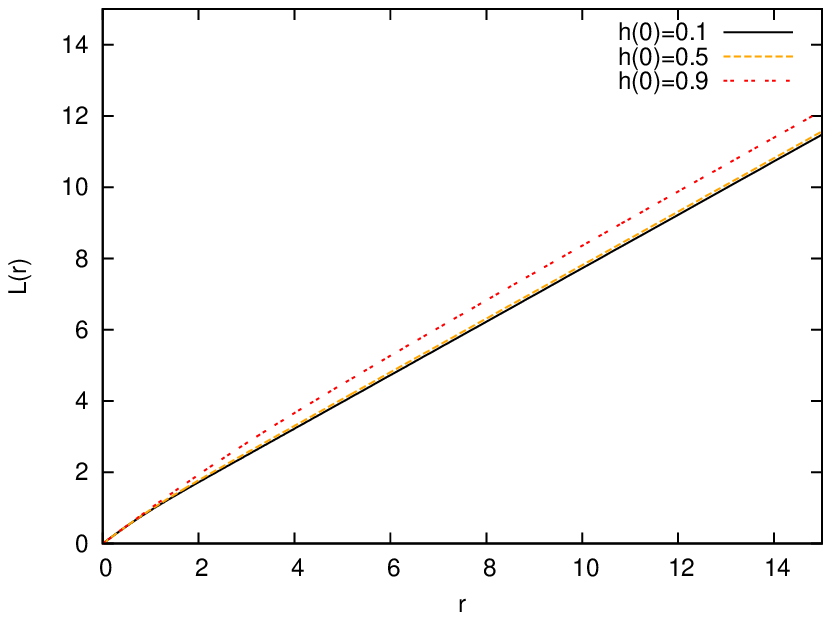}
\caption{\label{fig:semi3}The function L for different configurations. On the left, $h(0)=0.5$ while we vary the values of $\alpha$; on the right $\alpha=0.5$ while we vary the value of $h(0)$.}
\end{figure}

The figures show that the metric function $L$ does indeed depend on the value 
of the condensate. We can see the dependence more clearly in  the right plot of Fig.~\ref{fig:semi3} where we have displayed the metric function $L$ for different values of the condensate keeping constant the value of $\alpha$. In the left plot of  Fig.~\ref{fig:semi3} we show the profile function $L$ in different cases where we keep the value of the condensate constant and vary the value of the $\alpha$ parameter. Note that the deficit angle (\ref{deficit}) increases with increasing $\alpha$, and also that  as $\alpha$ approaches zero, the space-time becomes Minkowski, where $L(r)= r$. \\

\begin{figure}[!htb]
\includegraphics[width=6cm]{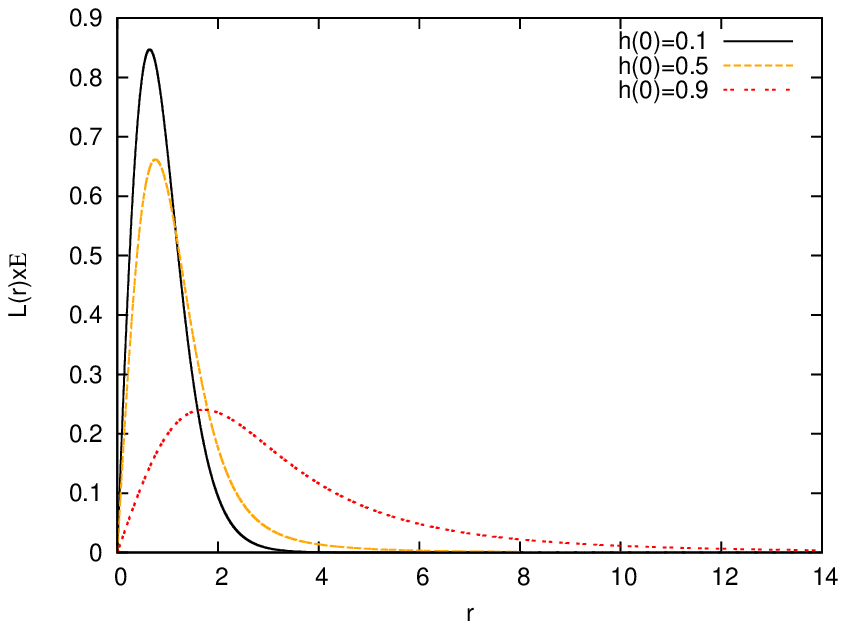} 
\caption{\label{fig:semi4}Energy density for different values of the free parameter $h(0)$, with fized $\alpha=0.5$.}
\end{figure}

One way of understanding this phenomenon is by looking at the energy density for each one of these configurations, as shown in Fig.~\ref{fig:semi4}, where we only vary the value of the condensate. 
Even though the total energy (the area below these lines) is the same, the energy density is  different, and affects the metric function with different strength. 
Indeed, since the width of the string core increases with the value of the condensate, the energy density spreads and the metric profile function $L$ reaches its asymptotic behavior farther away from the string. Note also that the deficit angle for each of the configurations is exactly the same,  and therefore the slope of the profile function $L$ in the limit $r\to \infty$ should be the same regardless of the value of the condensate. 

For BPS cosmic strings the energy density is closely related  to the  magnetic  field $F_{r \varphi}$
\be
\cE(r) = \xi \frac{v'(r)}{L},
\ee
and thus from (\ref{semilocalAtZero}) and (\ref{semilocalAtInfty}) it is trivial to find the limiting form of the energy density
\be
\cE(r\to 0) \approx \frac{\xi}{2} + \ldots \, , \qquad 
\cE (r\to \infty) \approx  \frac{2 \xi (|n|-|m|)^2}{(1-|n|\alpha^2)^2 r^2} \left( \frac{r}{r_0} \right)^{\frac{2(|m|-|n|)}{1 - |n|\alpha^2}}+ \ldots \, .
\ee

\begin{figure}[!htb]
\includegraphics[width=6cm]{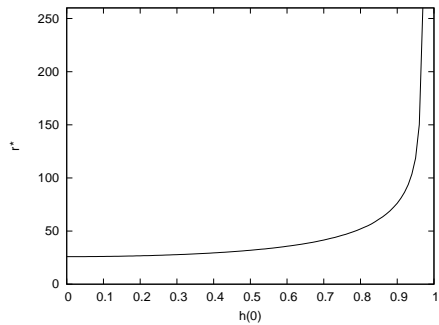} 
\caption{\label{fig:semi5}Values $r^*$  at which the function $L=0$ for a given coupling constant ($\alpha=1.02$) with respect to 
the value of the condensate at the core $h(0)$.}
\end{figure}

A dramatic effect happens when considering supermassive strings. In this case the string is massive enough to make  the function $L$ turn round and become zero at some finite value of $r^*$ (see Fig \ref{fig:semi3} for the case with $\alpha=1.02$), 
in other words, the deficit angle $\delta$ becomes larger than $2 \pi$. 
The value of the condensate in the core determines the extent to which this solution exists; it decides the ``size'' of the universe.  Once again, the total energy 
does not change, i.e.,  the integral for the energy density curve from $r=0$ up to the other point $r^*$ where $L(r^*)=0$  is independent of the value of 
$h(r=0)$. In Fig. \ref{fig:semi5} we have depicted the values of $r^*$  for a fixed coupling constant with respect to  the value of the condensate at the core $h(r=0)$. As the value of the condensate increases the string core (where space-time is approximately Minkowski) becomes wider. 
In the limiting case where $h(r=0)\to1$ the space-time becomes Minkowski everywhere and  the point $r^*$ tends to infinity.

\subsection{Tachyonic strings or $\phi$ strings}

These solutions share many properties with the semilocal strings, the main difference being the factor of $h^3$ in the equation 
(\ref{axionicBPSeq}) instead of the $h$ in equation (\ref{BPSprofiles}). This translates into the function $h$ tending to zero logarithmically. As in the semilocal case, 
the tachyonic field $f$ is responsible for the formation of the string, whereas $h$ is responsible for the condensate. Close to the core of the string the 
profile functions behave as in flat space-time \cite{BlancoPillado:2005xx}, in particular
\be
f(r) \approx f_0 \, r^{|n|} + \ldots \, , \qquad  h(r)^{-2} \approx h_0^{-2} -2 \alpha^2 q|m| \log r + \ldots \, , \qquad v(r) \approx \ft12 r^2 + \ldots \, .
\label{solAtCore}
\ee
  
\begin{figure}[!htb]
\includegraphics[width=6cm]{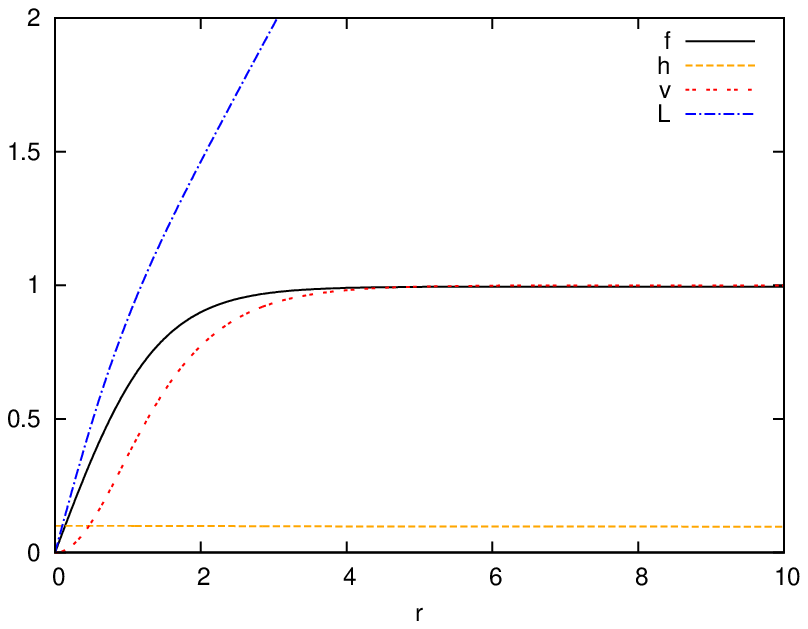}
\includegraphics[width=6cm]{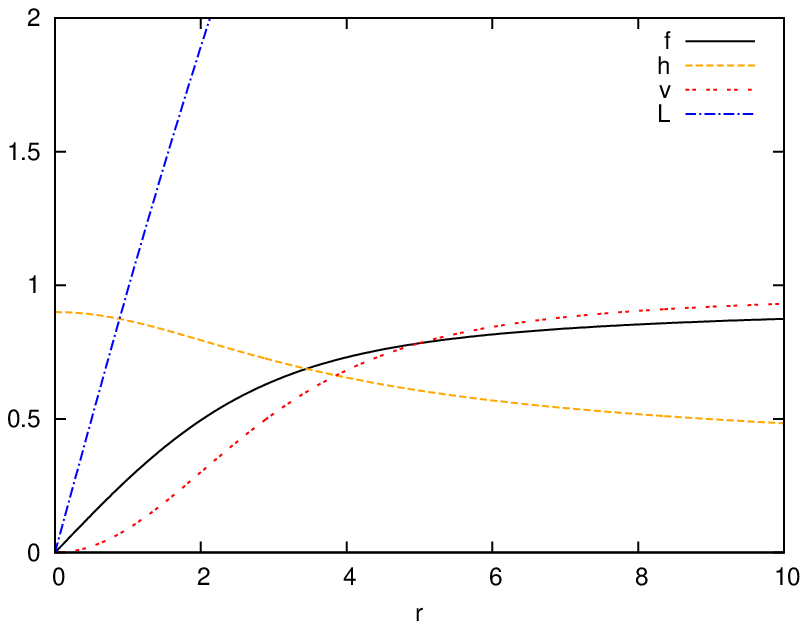} 
\caption{\label{fig:tac1}Profiles of a  gravitating tachyonic string  ($\alpha=0.5$)  with the value of the free parameter $h(0)=0.1$ (left) and
$h(0)=0.9$ (right).}
\end{figure}

The value $h(0)$ is a free parameter,
 related to the string width, which leaves the total energy unchanged and modifies the field configurations slightly. A typical $n=1$ cosmic string configuration can be seen in Fig. \ref{fig:tac1}, which shows that, as in the case of semilocal strings, for larger values of the condensate the core width increases. Note also how the $h$ field tends to zero slowly. Actually, all the profile functions approach their values at infinity logarithmically, as 
can be seen from their asymptotic expansion
\be
h(r)^2 \approx \frac{(1-|n|\alpha^2)}{2 \alpha^2 (|n|- q |m| ) \log r} + \ldots, \quad f(r ) \approx 1 - \frac{(1-|n|\alpha^2)}{4 \alpha^2 (|n|- q |m| ) \log r} +\ldots, 
\quad v(r) \approx \frac{|n|}{q} - \frac{(1-|n|\alpha^2)^2}{4 \alpha^2 q (|n|- q |m| ) \log^2 r}+\ldots .
\label{tachyonicAtInfty}
\ee
The term appearing in the numerator of the three expressions $(1-|n|\alpha^2)$ was not obtained in the flat space-time analysis done in \cite{BlancoPillado:2005xx}, 
since this is a consequence of the asymptotic form of the space-time metric
\be
ds^2 \approx - dt^2 + dr^2 + (1-|n|\alpha^2)^2 \, r^2 d \varphi^2 + dz^2 .
\ee
From (\ref{tachyonicAtInfty}) we can immediately extract the form of the energy density far away from the core:
\be
\cE(r) = \frac{M_p^2 (1 - |n| \alpha^2)}{2 (|n| - q |m|) r^2 \log^3 r } +\ldots \, .
\ee

\begin{figure}[!htb]
\includegraphics[width=6cm]{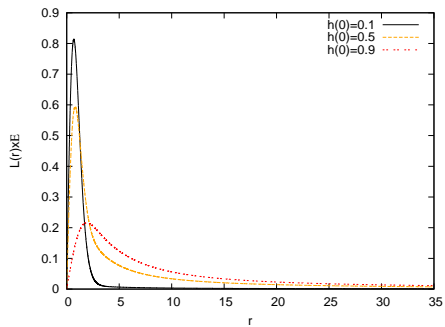} 
\caption{\label{fig:tac2} Energy density for different values of the free parameter $h(0)$ with fixed $\alpha=0.5$.}
\end{figure}

\begin{figure}[!htb]
\includegraphics[width=6cm]{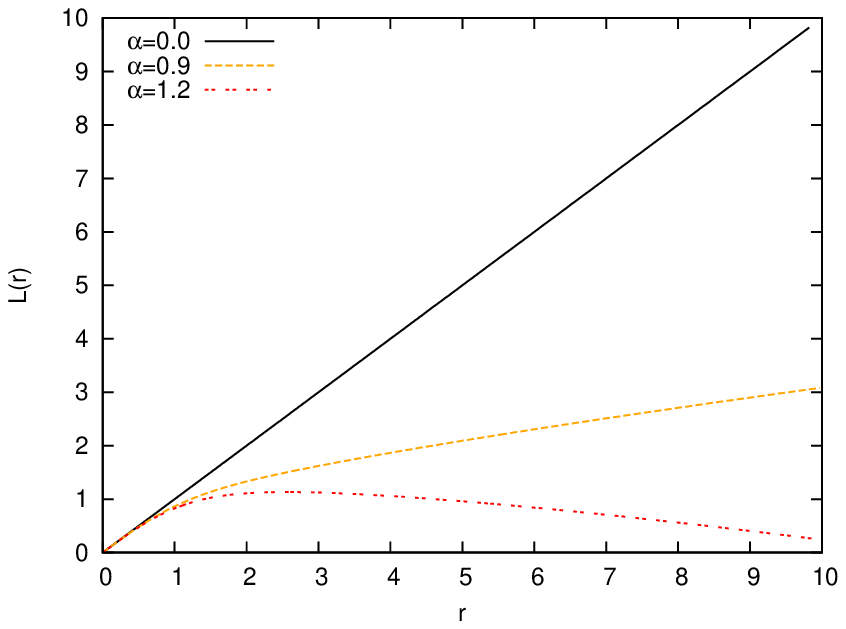} 
\includegraphics[width=6cm]{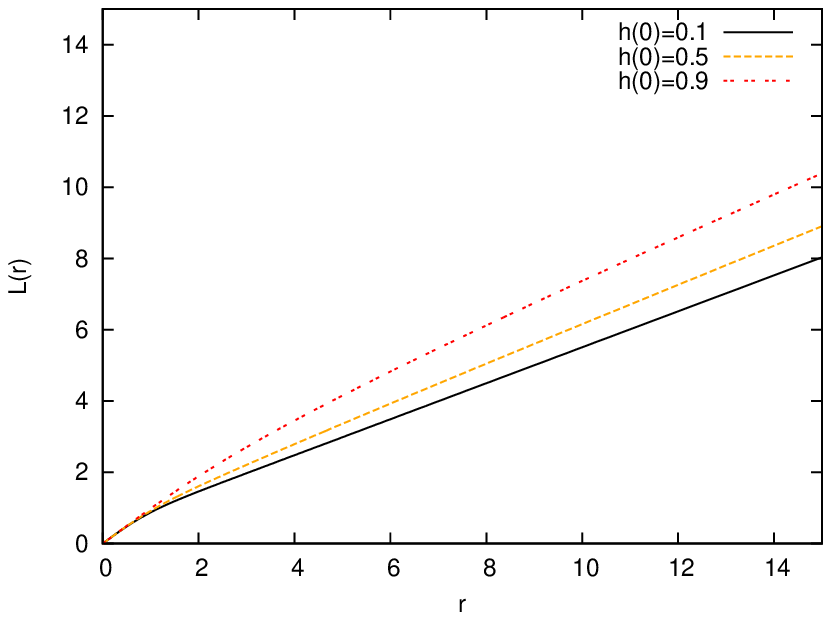}
\caption{\label{fig:tac31}The function L for different models: left, $h(0)=0.5$ and various $\alpha$; right, $\alpha=0.5$ and various $h(0)$. }
\end{figure}

Most of our results for the semilocal strings can also be obtained here. The global energy, and thus the deficit angle, are independent of the value of the $h$ field at $r=0$,  implying that the slope of $L$ remains unchanged for large values of $r$; however, as the energy density spreads due to the presence of the condensate,  the behaviour of the  metric profile function $L$ changes close to the core, and in particular reaches its asymptotic behavior farther away from the string center (see Fig.~\ref{fig:tac31}).

\begin{figure}[!htb]
\includegraphics[width=6cm]{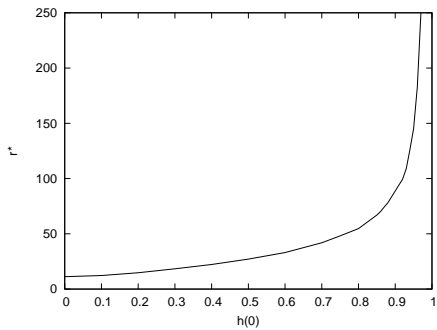} 
\caption{\label{fig:tac4}Values of the points $r^*$ at which the function $L=0$ for a given coupling constant $\alpha=1.1$ with respect to 
the value of the condensate at the core $h(0)$.}
\end{figure}

 Once again, the deficit angle increases for larger values of $\alpha$, and in the case of supermassive strings, the metric field develops a zero far from the core, making the space-time closed. Fig.~\ref{fig:tac4} 
depicts the points $r^*$ where the $L$ develops a second zero for different values of the condensate at the core,  keeping  the value of $\alpha$ fixed. 

\subsection{Axionic strings or $s$-strings}

\begin{figure}[!htb]
\includegraphics[width=6cm]{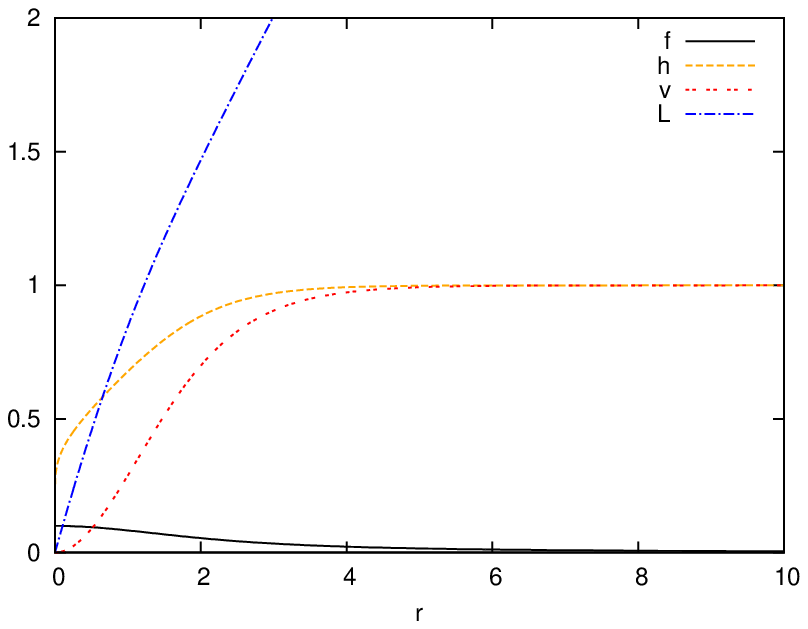}
\includegraphics[width=6cm]{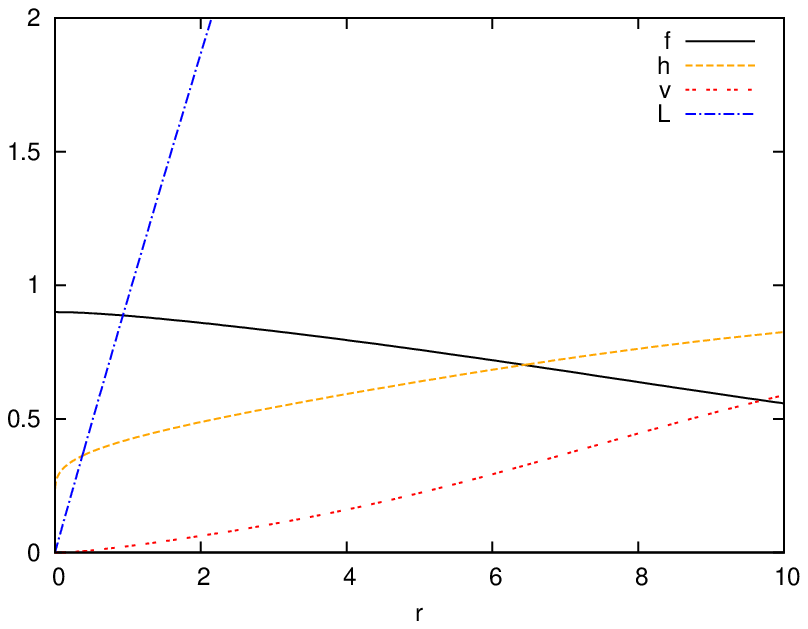} 
\caption{\label{ax1}The profiles of a  gravitating axionic string ($\alpha=0.5$)  with the value of the free parameter $f(0)=0.1$ (left) and
$f(0)=0.9$ (right).}
\end{figure}

For type of strings, the roles played by the tachyon field $f$ and the dilaton field $h$ are interchanged, the latter being responsible for 
the formation of the strings, and the former giving a measure of the width of the string (see Fig.~\ref{ax1}) . Close to the core, the approximate form of the profile functions is also given by (\ref{solAtCore}), but in this case $q |m|> |n|$, and in 
particular Fig.~\ref{ax1} corresponds to the case $n=0$.  Here $f(0)$ plays the role of the free parameter which fixes the width of the string. The following asymptotic expansion shows the behavior of the profile functions in the opposite 
limit, $r \to \infty$
\be
f(r) \approx \left(\frac{r}{r_0}\right)^{\frac{|n|-q|m|}{1-q|m|\alpha^2}}+\ldots, \qquad h(r)\approx 1 - \frac{1}{2} \,\left(\frac{r}{r_0}\right)^{\frac{2(|n|-q|m|)}{1-q|m|\alpha^2}}+\ldots, \qquad 
v(r) \approx |m| - \frac{|n|-q|m|}{q \alpha^2} \, \left(\frac{r}{r_0}\right)^{\frac{2(|n|-q|m|)}{1-q|m|\alpha^2}}+\ldots \,.
\label{axionicAtInfty}
\ee
This result is slightly different than the one obtained in a Minkowski background  \cite{BlancoPillado:2005xx}, as can be seen from the appearance of the factor $(1- q |m|\alpha)$ in the 
exponents. As for the case of the tachyonic strings, this correction is related to the form of the space-time metric far away from 
the core of the string:
\be
ds^2 \approx - dt^2 + dr^2 + (1-q|m|\alpha^2)^2 \, r^2 d \varphi^2 + dz^2 .
\ee
From (\ref{axionicAtInfty}) we obtain the asymptotic expansion of the energy density far away from the core:
\be
\cE(r) = \frac{2 M_p^2 (|n| - q |m|)^2 }{(1-q|m|\alpha^2)^2 r^2}\left(\frac{r}{r_0}\right)^{\frac{2(|n|-q|m|)}{1-q|m|\alpha^2}} +\ldots \, .
\ee

The energy density of these strings is close to that of the semilocal or tachyonic strings, in that there is a concentration of energy next to the core, although at much shorter distances than in those cases. This effect has  its origin in the factor $\Re(S)^{-2}$ of the kinetic terms of the axio-dilaton field (\ref{AxioDilatonEnergy}), 
which is responsible for the divergence of $\Re(S)$ at the center of the  $s-$strings. As in the two types of strings previously discussed, the details of the shape of the energy density depend on the value of the condensate. 

\begin{figure}[!htb]
\includegraphics[width=6cm]{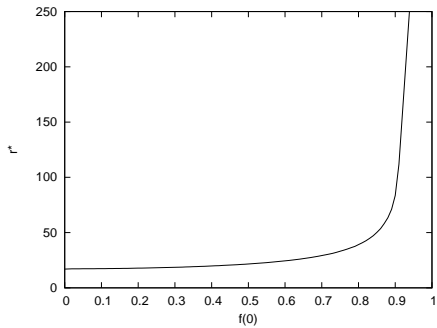}  
\caption{\label{ax3}Values of the points $r^*$ at which the function $L=0$ for a given coupling constant $\alpha=1.1$ with respect to 
the value of the condensate at the core $f(0)$.}
\end{figure}

Once again, the global energy of these configurations does not change with respect to the value of the condensate at the core, the zero mode 
survives. In this case,  solutions for supermassive strings that have a closed space-time associated with them can also  be obtained, and the dependency of the values $r^*$ at which the closing occurs can be found in Fig.~\ref{ax3}.

\section{Conclusions and discussion}
\label{conclusions}

In this paper we have studied field theoretical models for cosmic strings with flat directions in curved space-time. 
Specifically we have considered the effects of the gravity coupling on models with semilocal, axionic and tachyonic strings, which were known to have flat 
directions  in flat space-time. In this work we have focused on solutions of a single static cylindrically symmetric string with invariance under boosts along the axis of symmetry. Although the models studied  are very different in character, 
some of the solutions obtained are very similar.\\ 

The cosmic string solutions have been found using a BPS-type of argument, which consists in minimizing an energy functional 
appropriate for static cylindrically symmetric field configurations in a gravitational theory. The field configurations saturate the corresponding BPS bound provided the fields satisfy a set of first order differential equations, similar to those found for Minkowski space. Following \cite{Gibbons:1992gt} we have shown that, if these conditions are met, then the energy momentum tensor acquires a very simple form, which allows to find a first integral to the Einstein equations. This result is also interesting in the context of supergravity theories, where this first integral is known as the gravitino equation, and it is known to exist whenever the cosmic string configuration preserves a fraction of the supersymmetries. However, since the derivation we have presented here makes no reference to supersymmetry, our results show that cosmic string solutions saturating a BPS bound can satisfy a gravitino-type of equation, regardless of the  fraction of supersymmetries of the model that is broken by the string.\\  

 As in the case of the non-gravitational version of these three types of strings, the set of differential equations characterizing them admit a one-parameter family of solutions which are degenerate in energy. Therefore, we have proved that the 
zero mode survives the coupling to gravity in the three cases. 
We have found numerical solutions to the BPS equations for the three types of strings, paying special attention to the space-time metric, and in particular we have characterized its dependence on the value of the family-parameter and the string tension. \\

Even though  the free parameter does not change the total energy, and therefore, neither the deficit angle 
of the string, it changes the shape of the energy density. This is due to the fact that  the free 
parameter changes the profiles of the fields forming the string, in particular it changes the width of the 
string core, which in turns  modifies the energy distribution. 
As gravity depends on the energy density, not only on the global value of it, 
the energy distribution changes the metric function $L$ 
which, the wider the string, the farther from the string center it reaches its asymptotic behavior. 
Thus, we show that different values of the zero mode do change the metric properties, even 
though they do not change the global characteristics, such as the deficit angle. 
This effect becomes very apparent when  the string is massive enough to have  a deficit angle larger than $2 \pi$. 
In those cases the $L$ function turns round and becomes zero at some finite value of the radial coordinate, 
rendering the spatial directions transverse to the string closed. As the free parameter associated with 
the zero mode changes the shape of the $L$ function,  its value acts as a modulus which fixes the 
size of the compact space-time dimension.\\

We would like to mention possible generalizations of this work. On the one hand, it would  be interesting to study whether the zero-mode associated with the string width would survive  the coupling to gravity in general situations, \emph{i.e.} situations involving  several strings with a random motion.  It is known that the semilocal model admits static parallel muti-vortex  solutions  with flat directions which survive the coupling to gravity \cite{Gibbons:1992gt}, and the low energy dynamics of the corresponding zero-modes has been studied in flat space-time \cite{Leese:1992fn}. Given the similarities between the semilocal strings and the axionic and tachyonic strings we expect that  performing  similar analyses for the later cases is possible, and in particular to find static multi-vortex solutions with flat directions even after including the gravity coupling.         
On the other hand, the study of the models in this paper was performed statically. The 
analysis of the dynamics of these models would also be very interesting, for example to investigate whether the zero modes could be excited, or whether the zero modes could actually be dynamical.  

\begin{acknowledgments}

We are grateful to J.J. Blanco-Pillado for very useful discussions.  BH and KS gratefully acknowledge support within the framework of the Deutsche Forschungsgemeinschaft (DFG) Research Training Group 1620 {\it Models of gravity}. JU acknowledges financial support from the Basque Government (IT-559-10), the Spanish Ministry (FPA2009-10612), and the Spanish Consolider-Ingenio 2010 Programme CPAN (CSD2007-00042). KS thanks the department of Theoretical Physics and Science History at the University of the Basque Country  for its hospitality.
KS was supported by DFG grant HA-4426/5-1.

\end{acknowledgments}

\bibliography{gravflat}

\end{document}